\documentclass[fleqn,12pt,twoside]{article}
\usepackage[headings]{espcrc1}

\usepackage{graphicx}
\usepackage{epsfig}
\usepackage{amssymb}
\usepackage{subfigure}
\usepackage{wrapfig} 
\usepackage{caption}

\usepackage[figuresright]{rotating}

\newcommand{\AmS}{{\protect\the\textfont2
  A\kern-.1667em\lower.5ex\hbox{M}\kern-.125emS}}

\usepackage{amsmath}

\title{$\phi$ production as seen in $e^+e^-$ and $K^+K^-$ decay channels in Au+Au
  collisions by PHENIX at $\sqrt{s_{NN}}$ = 200 GeV} 

\author{ A. Kozlov \address[WIS]{Weizmann Intitute of Science, Rehovot 76100, Israel} for the PHENIX
  collaboration\thanks{for the full list of PHENIX authors and acknowledgements, see
    Appendix "Collaborations" of this volume.} }

\begin{document}

\maketitle

\begin{abstract}
  The properties of the $\phi$-meson have been measured via its $e^+e^-$ and $K^+K^-$
  decay channels in Au+Au collisions at $\sqrt{s_{NN}}$ = 200 GeV by the PHENIX experiment. The
  preliminary yields and temperatures derived for the minimum bias and several centrality bins in both
  decay channels are presented.
  
\end{abstract}

\section{Introduction}
The $\phi$-meson is an important diagnostic tool for the state of matter produced in relativistic 
heavy ion collisions. It can be sensitive to chiral symmetry restoration through in-medium
modification of its spectral properties (mass and/or shape)~\cite{Pal01}. In particular, since
$m_{\phi} \approx 2 m_{K}$, even small changes in the spectral properties of the $\phi$ or $K$ can
induce significant changes in the $\phi \rightarrow 
K^+K^-$ yield. The PHENIX detector has an excellent mass resolution and is capable of measuring both
$e^+e^-$ and $K^+K^-$ decay channels simultaneously. The present results are from the Au+Au
collisions at $\sqrt{s_{NN}}$ = 200 GeV measured by PHENIX during the 2004 RHIC run.

\section{PHENIX experiment}

The analysis was performed using the two central arms of the PHENIX spectrometer~\cite{Adc01} which
cover the pseudorapidity range $\vert\eta\vert <$ 0.35 and 2 $\times$ 90$^{\circ}$ in
azimuthal angle $\phi$. The momentum and charge were determined using the drift chamber (DC) and the
pad chamber (PC1). Valid
DC-PC1 tracks were confirmed by the matching of the projected and associated hit information
to the Ring Imaging Cherenkov (RICH) detector and  electromagnetic calorimeter (EMCal) in the case of
electrons and to the Time Of Flight (TOF) detector or EMCal in the case of kaons.

Electrons were identified by the RICH and by requiring the energy in the EMCal to match the measured
momentum of the tracks. 
Kaons were identified using the timing information from the TOF and EMCal which have very good
$\pi$/K separation in the  
momentum range 0.3 $< p(GeV/c) <$ 2.5 and 0.3 $< p(GeV/c) <$ 1.0, respectively.

The beam-beam counters (BBC) and the zero-degree calorimeters (ZDC) provided the trigger and
determined the event centrality. The BBC's were also used to determine the $z$-coordinate of the
collision vertex ($z_{vtx}$).

The $\phi \rightarrow e^+e^-$ analysis used 903 $\times$ 10$^6$ minimum bias events with a vertex
position within -~28~$ < z_{vtx} (cm) <$~26. Kaons identified by TOF and 
EMCal were combined in pairs using four different detector combinations: TOF-TOF, TOF-EMC$_{EAST}$,
EMC$_{EAST}$-EMC$_{EAST}$, EMC$_{WEST}$-EMC$_{WEST}$. The analysis performed with TOF-TOF (other
detector combinations) used 409 (170) $\times$ 10$^6$ minimum bias events with a vertex
position within $\vert z_{vtx} (cm) \vert < $ 30.

\section{Analysis procedure}

The $\phi$ meson yield (dN/dy) and temperature (T) were derived from the invariant
m$_T$-distribution: 

\begin{equation}
  \frac{1}{2 \pi m_{T}} \frac{d^{2}N}{dm_{T}dy}  = \frac{N^{\phi}_{raw}(m_{T}){\cdot}CF(m_T)}
  {2\pi m_T{\cdot}N_{events}{\cdot}\epsilon_{pair-embedding}{\cdot}\epsilon_{run-by-run}{\cdot}BR{\cdot}\Delta m_{T}},
\end{equation}

where $N^{\phi}_{raw}(m_{T})$ is the raw $\phi$ yield,  $CF(m_T)$ is the correction factor to
account for acceptance and pair reconstruction efficiency, $N_{events}$ is the number of analyzed
events, $\epsilon_{pair-embedding}$ is the pair embedding efficiency to account for the
reconstruction efficiency losses due to detector occupancy, $\epsilon_{run-by-run}$ is an efficiency
which takes into account run-by-run variations of the detector performance, $BR$ is the branching
ratio for $\phi \rightarrow e^+e^-$ or $K^+K^-$ and $\Delta m_{T}$ is the bin size.

The raw $\phi$ yield for each decay channel was determined in two steps. First, all identified
particles in a given event were   
combined in pairs to form like- and unlike-sign invariant mass spectra. The inherent
combinatorial background was estimated by using an event mixing procedure in which the particles 
from one event were combined with the particles from the next twenty events provided
that all events belong to the same centrality and vertex classes. The unlike-sign mixed event
integral yield was normalized to the measured 2~$\sqrt{N^{++}N^{--}}$ yield. The validity of this method
of event mixing was confirmed by comparing the like-sign invariant mass spectra from mixed events to
the measured one. In the second step, we subtracted the normalized unlike-sign mixed event spectrum 
from the measured one and derived the mass distribution. In the dielectron channel the $\phi$ meson
yield derived for the mass window 0.997 $< m (GeV/c^2) <$ 1.041 was about 900 counts with a signal
to background (S/B) ratio of about 1/50. The $\phi$ meson yield in kaon channel (mass window  1.014 $< m
(GeV/c^2) <$ 1.024) was about 44000 counts with S/B of $\sim$ 1/10.

The correction factor $CF(m_T)$ was determined using Monte Carlo simulations. Single
$\phi$-mesons were generated with an exponential transverse momentum distribution.
The $\phi$'s were decayed, propagated  through an emulator of the PHENIX detector and the resulting
output was passed through the whole analysis chain. For each m$_T$ bin the ratio of the generated
yield to the reconstructed one gives the correction factor $CF(m_T)$.

Finally, the dN/dy and T were extracted from the corrected invariant m$_T$ spectra fitted with the
following exponential function  having dN/dy and T as parameters:
\begin{equation}
  \label{eq:eq_fit}
    \dfrac{1}{2\pi m_{T}}\dfrac{d^{2}N}{d m_{T} dy} = \dfrac{dN/dy}{2
      \pi T (T+M_{\phi})} exp(-(m_{T} - M_{\phi})/T),
\end{equation}

where M$_{\phi}$ is the PDG value of the $\phi$ meson's mass, 1.019 GeV/c$^2$.

\section{Results}

\begin{figure}[h]
  \vspace{-7mm}
  \begin{center}
    \includegraphics[width=0.42\linewidth]{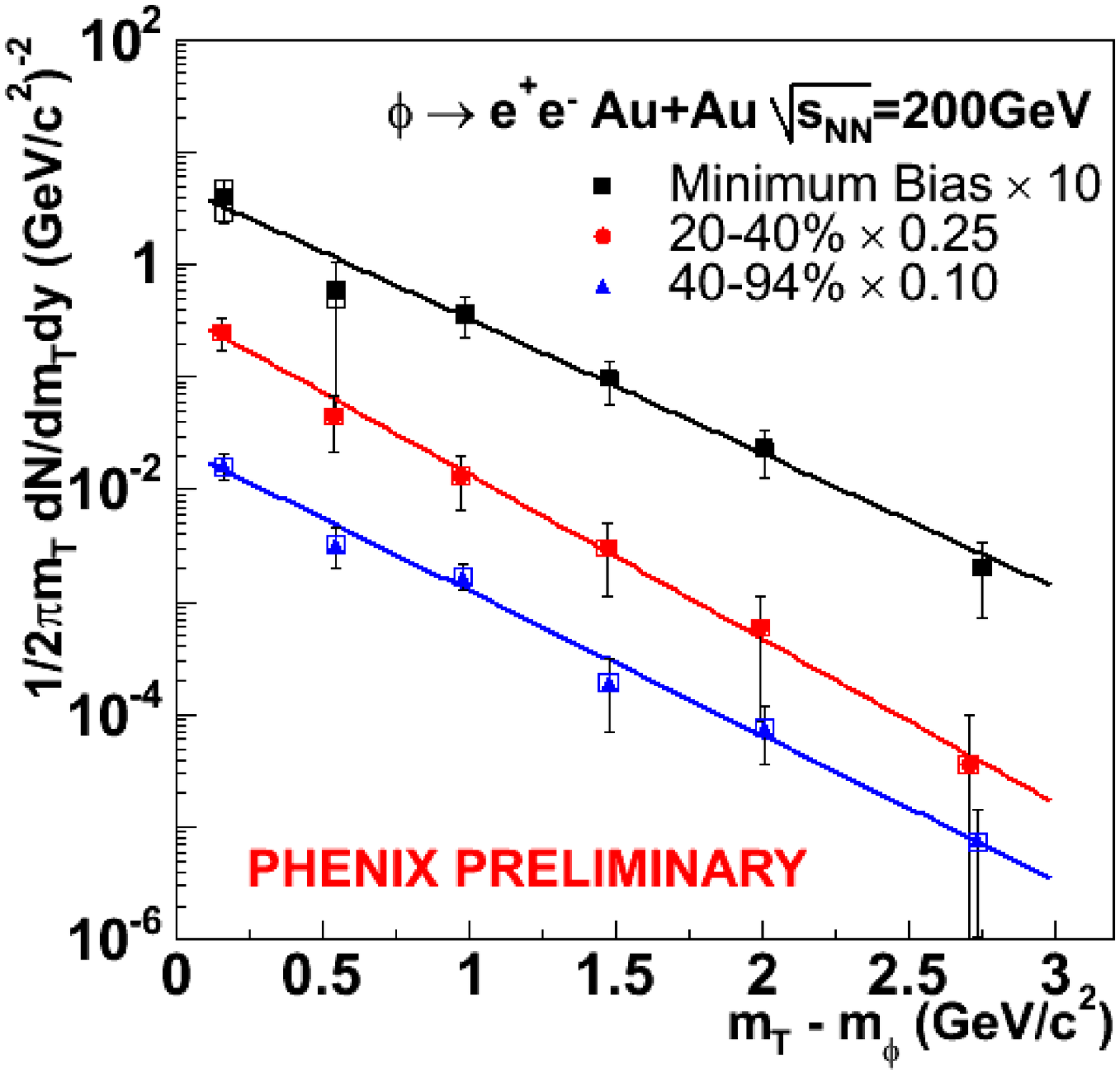}
    \hfill
    \includegraphics[width=0.41\linewidth]{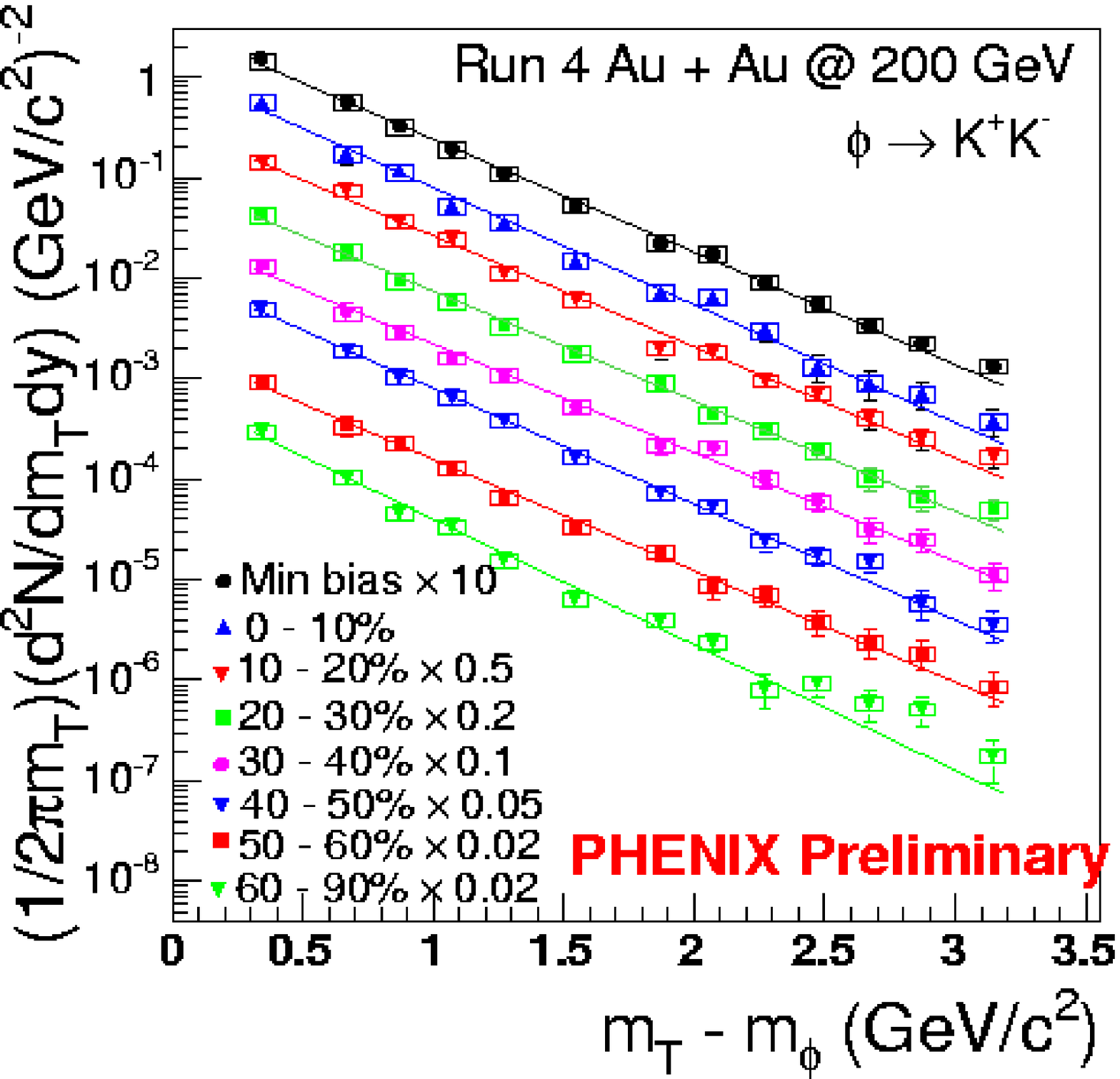}
  \end{center}
  \vspace{-15mm}
  \caption{Invariant m$_T$ spectra. Statistical and systematic errors are shown by vertical bars and
    open rectangles, respectively.} 
  \label{fig:fig_invt} 
  \vspace{-7mm}
\end{figure}

In this contribution we present only m$_T$ spectra, yield and temperature results derived from the
measurements of the $\phi$-meson through the two decay channels. The invariant m$_T$ spectra for
minimum bias and several centrality bins fitted with the exponential function Eq.~\ref{eq:eq_fit}
are shown in the left and right panels of Fig.~\ref{fig:fig_invt} for $\phi \rightarrow e^+e^-$
and $\phi \rightarrow K^+K^-$, respectively. 

The yields and temperatures extracted from the fits are summarized in Table~\ref{tab:tab_dndy_t_ee}
and Table~\ref{tab:tab_dndy_t_kk}. The systematic errors quoted in the tables result from the
uncertainties both in the simulation and in the analysis procedure. The dN/dy and T obtained in the $K^+K^-$
measurements are in agreement with the results of RHIC run 2002~\cite{DD01} within the statistical
and systematic errors.  

\begin{table*}[htp]
  \vspace{-5mm}
  \caption{Summary of dN/dy and T for the $\phi \rightarrow e^+e^-$ analysis.}
  \label{tab:tab_dndy_t_ee}
  \newcommand{\m}{\hphantom{$-$}}
  \newcommand{\cc}[1]{\multicolumn{1}{c}{#1}}
  \begin{center}
    \begin{tabular}{ccc}
      \hline 
      \textbf{Centrality \%} &    dN/dy    & T (GeV) \\
      \hline                                                                 
      MB      &  1.54 $\pm$ 0.43(stat) $\pm$ 0.43(syst) & 0.364 $\pm$ 0.031(stat) $\pm$ 0.025(syst)    \\
      20 - 40 &  3.64 $\pm$ 0.89(stat) $\pm$ 0.76(syst) & 0.297 $\pm$ 0.036(stat) $\pm$ 0.033(syst)   \\
      40 - 94 &  0.66 $\pm$ 0.12(stat) $\pm$ 0.12(syst) & 0.338 $\pm$ 0.030(stat) $\pm$ 0.041(syst)   \\
      \hline 
    \end{tabular}
  \end{center}
  \vspace{-10mm}
\end{table*}

\begin{table*}[htp]
  \vspace{-7mm}
  \caption{Summary of dN/dy and T for the $\phi \rightarrow K^+K^-$ analysis.}
  \label{tab:tab_dndy_t_kk}
  \newcommand{\m}{\hphantom{$-$}}
  \newcommand{\cc}[1]{\multicolumn{1}{c}{#1}}
  \begin{center}
    \begin{tabular}{ccc}
      \hline 
      \textbf{Centrality \%} &    dN/dy    & T (GeV) \\
      \hline                                                                 
      MB      &  1.08 $\pm$ 0.04(stat) $\pm$ 0.20(syst) & 0.388 $\pm$ 0.005(stat) $\pm$ 0.027(syst)    \\
       0 - 10 &  3.80 $\pm$ 0.30(stat) $\pm$ 0.72(syst) & 0.372 $\pm$ 0.011(stat) $\pm$ 0.026(syst)   \\
      10 - 20 &  2.32 $\pm$ 0.16(stat) $\pm$ 0.44(syst) & 0.394 $\pm$ 0.010(stat) $\pm$ 0.027(syst)   \\
      20 - 30 &  1.62 $\pm$ 0.11(stat) $\pm$ 0.31(syst) & 0.397 $\pm$ 0.010(stat) $\pm$ 0.028(syst)   \\
      30 - 40 &  0.95 $\pm$ 0.07(stat) $\pm$ 0.18(syst) & 0.401 $\pm$ 0.010(stat) $\pm$ 0.028(syst)   \\
      40 - 50 &  0.75 $\pm$ 0.04(stat) $\pm$ 0.13(syst) & 0.377 $\pm$ 0.008(stat) $\pm$ 0.026(syst)   \\
      50 - 60 &  0.35 $\pm$ 0.03(stat) $\pm$ 0.06(syst) & 0.392 $\pm$ 0.012(stat) $\pm$ 0.027(syst)   \\
      60 - 90 &  0.11 $\pm$ 0.01(stat) $\pm$ 0.02(syst) & 0.348 $\pm$ 0.011(stat) $\pm$ 0.024(syst)   \\
      \hline 
    \end{tabular}
  \end{center}
  \vspace{-7mm}
\end{table*}

\begin{figure}[h]
  \vspace{-10mm}
  \begin{center}
    \includegraphics[width=0.45\linewidth]{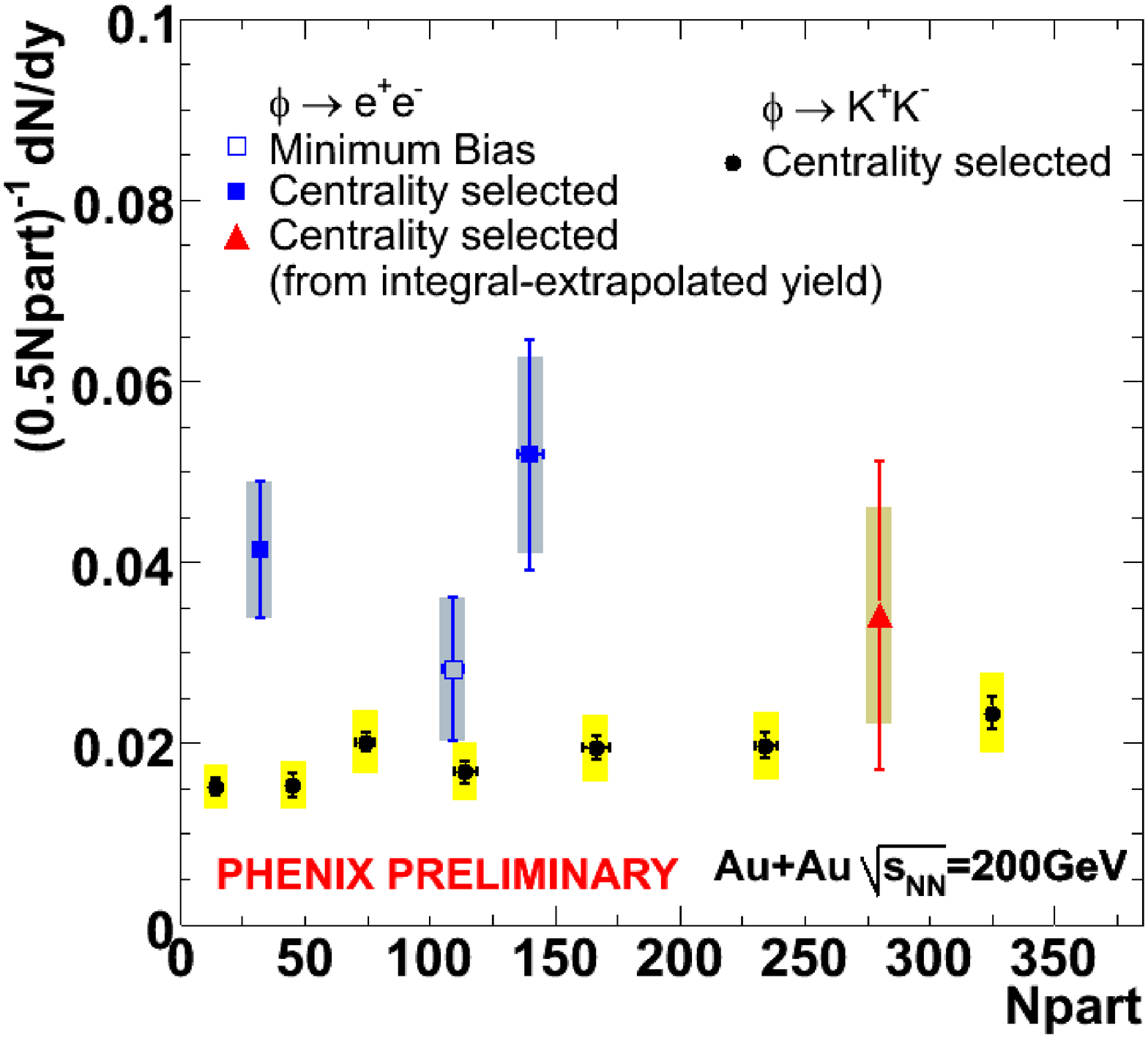}
    \hfill
    \includegraphics[width=0.45\linewidth]{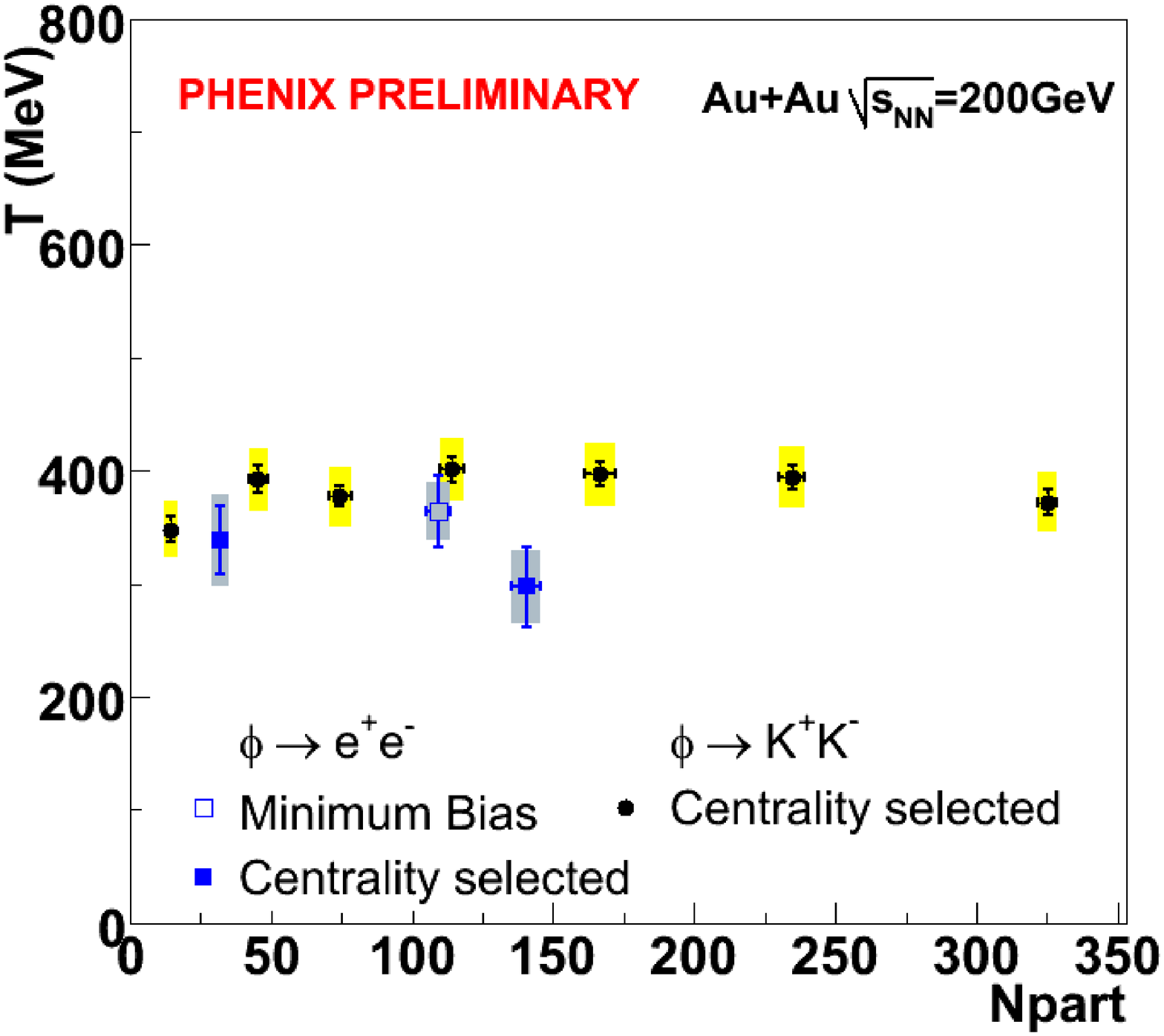}
  \end{center}
  \vspace{-15mm}
  \caption{Multiplicity dependence of the yield dN/dy (left panel) and temperature T (right panel)
    for $e^+e^-$ and $K^+K^-$ decay channels. Open and filled symbols in the dN/dy and T plots represent MB and
    centrality selected events, respectively. The triangle shows the $\phi$ yield derived from an independent
    analysis. Statistical and systematic errors are shown by vertical bars and shaded bands, respectively.}
    \label{fig:fig_dndy_temp} 
    \vspace{-10mm}
\end{figure}

Fig.~\ref{fig:fig_dndy_temp} shows the comparison of yields dN/dy (left panel) and  temperatures T
(right panel) extracted in the
two decay channels. The dielectron yield in the highest centrality bin has limited statistics and
its dN/dy (shown in the left panel of Fig.~\ref{fig:fig_dndy_temp} as a triangle) was derived from
an independent analysis using the integral yield. The corresponding correction factor integrated
over m$_T$ was obtained under the assumption that T = 366 MeV.

The temperatures measured in the $e^+e^-$ and $K^+K^-$ decay channels are in
agreement within errors.  The dN/dy value in the $e^+e^-$ decay channel seems to be larger than in the 
$K^+K^-$ decay channel. However, although the present data quality is significantly improved with
respect to the run-2 results~\cite{DD01,DD02}, the errors in the dielectron channel, both
statistical and systematic, are still too large for a definite statement. Further improvement in the
measurements of the dielectrons will be achieved with the planned upgrade of the PHENIX detector
with a novel Hadron Blind Detector (HBD) which is expected to significantly reduce the combinatorial
background~\cite{IR01}. 

\end{document}